\font\tenfrakturb=eufb10
\font\tenfraktur=eufm10
\font\tenmsbm=msbm10
\font\sevenfrakturb=eufb7
\font\sevenfraktur=eufm7
\font\sevenmsbm=msbm7
\font\fivefrakturb=eufb5
\font\fivefraktur=eufm5
\font\fivemsbm=msbm5
\newfam\bgothicfam
\newfam\gothicfam
\newfam\msbmfam
\textfont\bgothicfam = \tenfrakturb \scriptfont\bgothicfam=\sevenfrakturb
\scriptscriptfont\bgothicfam=\fivefrakturb
\textfont\gothicfam = \tenfraktur \scriptfont\gothicfam=\sevenfraktur
\scriptscriptfont\gothicfam=\fivefraktur
\textfont\msbmfam = \tenmsbm \scriptfont\msbmfam=\sevenmsbm
\scriptscriptfont\msbmfam=\fivemsbm

\def\Bbb{\tenmsbm\fam\msbmfam}

\catcode`@=11
\def\renewcounter#1{\@definecounter{#1}\@ifnextchar[{\@newctr{#1}}{}}

\documentclass[10pt,a4paper]{article} 

\begin{document}

\vglue 1 truecm

\vbox{ {\em Europhys. Lett.}, {\bf 62}(5), pp. 684-690 (2003)
\hfill 1 June 2003
}
  
\vfil
\centerline{\large\bf Dirac equation in the confining
SU(3)-Yang-Mills field}
\vskip.5cm
\centerline{\large\bf and relativistic effects in charmonium spectrum}
  
\bigskip
\centerline{ Yu. P. Goncharov } 
\vspace{1true cm} 
\centerline{\em Theoretical Group, Experimental Physics Department,} 
\vskip.5cm
\centerline{\em State Polytechnical University, Sankt-Petersburg 195251, 
Russia}
  
\vfil
\begin{abstract}
The recently obtained solutions of the Dirac equation in the confining
SU(3)-Yang-Mills field in Minkowski spacetime are applied to describe
the energy spectrum of charmonium. The nonrelativistic 
limit is considered for the relativistic effects to be estimated in a 
self-consistent way and it is shown that the given effects could be extremely 
important for both the energy spectrum and the confinement mechanism.
\end{abstract}

\vfil
\hrule width 5truecm
\vskip .2truecm
\begin{quote} 
PACS: 
\hbox{12.38.-t} -- {Quantum chromodynamics.}
\hbox{12.38.Lg} -- {Other nonperturbative calculations.}
\hbox{14.40.Lb} -- {Charmed mesons.}
\end{quote}

\section{Introduction}
  Theory of quarkonium ranks high within hadron physics as the one of
central sources of information about the quark interaction. Referring for
more details to the recent up-to-date review\cite{Grin00}, it should be
noted here that at present some generally accepted relativistic model of
quarkonium is absent. The description of quarkonium is actually implemented
by nonrelativistic manner on the basis of the Schr{\"o}dinger equation
(concerning the general ideology here see, {\it e. g.}, ref. \cite{GM})
and then one tries to include relativistic corrections in one or another way.
Such an inclusion is not single-valued and varies in dependence of the point
of view for different authors (see, {\it e. g.}, ref.\cite{Du00} and references
therein). It would be more consistent, to our mind, building a primordially
relativistic model so that one can then pass on to the nonrelativistic one by
the standard limiting transition and, thus, to estimate the relativistic
effects in a self-consistent way.

 As follows from the main principles of quantum chromodynamics (QCD),
the suitable relativistic models for a description of the relativistic bound 
states of quarkonium could consist in considering the solutions of the Dirac 
equation in a SU(3)-Yang-Mills field representing the gluonic field. Indeed,
the Dirac equation in a SU(3)-Yang-Mills field is the direct consequence of 
the QCD Lagrangian in the same way as the Dirac equation for the hydrogen atom
is the direct consequence of the quantum electrodynamics (QED) Lagrangian.
Following the latter analogy, the mentioned SU(3)-Yang-Mills field should be
the so-called confining solution of the corresponding Yang-Mills equations,
{\it i. e.}, it should model the quark confinement. Such solutions are usually 
supposed to contain at least one
component of the mentioned SU(3)-field linear in $r$, the distance between
quarks. Recently, in Ref.\cite{Gon01} a number of such solutions have been
obtained and the corresponding spectrum of the Dirac equation describing the
relativistic bound states in those confining SU(3)-Yang-Mills fields
has been analysed. In this paper we should like to apply the results of
ref. \cite{Gon01} to the description of the charmonium spectrum.
Here we solve the inverse problem, {\it i. e.}, we define the confining gluonic 
field components in the covariant description (SU(3)-connection) for charmonium 
(which corresponds to a potential of $q\overline{q}$-interaction in the
nonrelativistic description) employing the experimental data on the mentioned 
spectrum\cite{pdg}. 
As a consequence, we shall not use any nonrelativistic potentials modelling 
confinement, for example, of the harmonic oscillator or funnel types, in 
particular, because the latter do not satisfy the Yang-Mills equations,
while the SU(3)-gluonic field used by us does. Accordingly, in our case the 
approach is relativistic from the very outset and our considerations are 
essentially nonperturbative, since we shall not use any expansions in the 
coupling constant $g$ or in any other parameters.

Further we shall deal with the metric of
the flat Minkowski spacetime $M$ that
we write down (using the ordinary set of local spherical coordinates
$r,\vartheta,\varphi$ for the spatial part) in the form
$$ds^2=g_{\mu\nu}dx^\mu\otimes dx^\nu\equiv
dt^2-dr^2-r^2(d\vartheta^2+\sin^2\vartheta d\varphi^2)\>. \eqno(1)$$
Besides, we have $|\delta|=|\det(g_{\mu\nu})|=(r^2\sin\vartheta)^2$
and $0\leq r<\infty$, $0\leq\vartheta<\pi$,
$0\leq\varphi<2\pi$.

  Throughout the paper we employ the system of units with $\hbar=c=1$,
unless explicitly stated otherwise.
Finally, we shall denote $L_2(F)$ the set of the modulo square integrable
complex functions on any manifold $F$ furnished with an integration measure
while $L^n_2(F)$ will be the $n$-fold direct product of $L_2(F)$
endowed with the obvious scalar product.

\section{Preliminaries}

  To formulate the results of ref.\cite{Gon01} that we need here, let us
notice that the relativistic wave function of the quarkonium can be chosen
in the form
$\psi=(\psi_1, \psi_2, psi_3)$
with the four-dimensional spinors $\psi_j$ representing the $j$-th colour
component of the quarkonium. The corresponding Dirac equation
for $\psi$ may look as follows:
$${\cal D}\psi=\mu_0\psi,\>\eqno(2)$$
where $\mu_0$ is a mass parameter and one can consider it to be the reduced
relativistic mass which is equal, {\it e. g.} for quarkonia, to one half the 
current mass of quarks forming a quarkonium,
while the coordinate $r$ stands for the distance between quarks.

From general considerations the explicit form of
the operator ${\cal D}$ in local coordinates $x^\mu$ on Minkowski
manifold can be written as follows:
$${\cal D}=i(\gamma^e\otimes I_3)E_e^\mu(\partial_\mu\otimes I_3
-\frac{1}{2}\omega_{\mu ab}\gamma^a\gamma^b\otimes I_3-igA_\mu),
\>a < b ,\>\eqno(3)$$
where $A=A_\mu dx^\mu$, $A_\mu=A^c_\mu T_c$ is a SU(3)-connection in the
(trivial) three-dimensional bundle $\xi$ over the Minkowski spacetime, 
$I_3$ is the unit matrix $3\times3$, the matrices $T_c$ form
a basis of the Lie algebra of SU(3) in the 3-dimensional space (we consider
$T_a$ to be Hermitian which is acceptable in physics), $c=1,...,8$, 
$\otimes$ here means tensorial product of matrices, $g$ is a gauge coupling 
constant. At last the coefficients $E_e^\mu$ and $\omega_{\mu ab}$ depend on
the choice of metric and their form for metric (1) can be found in 
ref. \cite{Gon01} as well as the explicit presentation of matrices 
$\gamma^a$, $a=0,...,3$.
As for the connection $A_\mu$ in bundle $\xi$, then, the suitable one should be
the confining solution of the Yang-Mills equations
$$d\ast F= \ast F\wedge A - A\wedge\ast F \>\eqno(4)$$
with the exterior differential $d=\partial_t dt+\partial_r dr+
\partial_\vartheta d\vartheta+\partial_\varphi d\varphi$ in coordinates
$t,r,\vartheta,\varphi$ while the curvature matrix (field strentgh)
for $\xi$-bundle is $F=dA+A\wedge A$ and $\ast$ means the Hodge star
operator conforming to metric (1).

 In ref.\cite{Gon01} the black hole physics techniques from
refs.\cite{Gon678} were used
to find a set of the
confining solutions of eq. (4).  For the aims of the given paper we need
one of these solutions of ref.\cite{Gon01}. Let us adduce it here
putting $T_c=\lambda_c$,
where $\lambda_c$ are the Gell-Mann matrices (whose explicit form can be
found in refs.\cite{Gon678}). Then the solution in question is the following
one:
$$ A^3_t+\frac{1}{\sqrt{3}}A^8_t =-\frac{a_1}{r}+A_1 \>,
 -A^3_t+\frac{1}{\sqrt{3}}A^8_t=\frac{a_1+a_2}{r}-(A_1+A_2)\>,
-\frac{2}{\sqrt{3}}A^8_t=-\frac{a_2}{r}+A_2\>, $$
$$ A^3_\varphi+\frac{1}{\sqrt{3}}A^8_\varphi =b_1r+B_1 \>,
 -A^3_\varphi+\frac{1}{\sqrt{3}}A^8_\varphi=-(b_1+b_2)r-(B_1+B_2)\>,
-\frac{2}{\sqrt{3}}A^8_\varphi=b_2r+B_2\> \eqno(5)$$
with all other $A^c_\mu=0$, where real constants $a_j, A_j, b_j, B_j$
parametrize the solution, and we wrote down
the solution in the combinations that are just
needed to insert into (2). From the adduced form it is clear that the solution 
is a configuration describing the electric Coulomb-like colour field 
(components $A_t$) and the magnetic colour field linear in $r$ (components 
$A_\varphi$).
Also, it is easy to check that the given solution satisfy the Lorentz gauge
condition that can be
written in the form ${\rm div}(A)=0$, where the divergence of the Lie algebra
valued 1-form $A=A^c_\mu T_cdx^\mu$ is defined by the relation
$${\rm div}(A)=\frac{1}{\sqrt{|\delta|}}\partial_\mu(\sqrt{|\delta|}g^{\mu\nu}
A_\nu)\>.\eqno(6)$$

 As was shown in ref.\cite{Gon01}, after inserting the above confining
solution into Eq. (2), the latter admits the solutions of the form
$$\psi_j=e^{i\omega_j t}r^{-1}\pmatrix{F_{j1}(r)\Phi_j(\vartheta,\varphi)\cr\
F_{j2}(r)\sigma_1\Phi_j(\vartheta,\varphi)}\>,j=1,2,3\eqno(7)$$
with the 2D eigenspinor $\Phi_j=\pmatrix{\Phi_{j1}\cr\Phi_{j2}}$ of the
Euclidean Dirac operator on the unit sphere ${\Bbb S}^2$.
The explicit form of $\Phi_j$ is not needed here and
can be found in
refs.\cite{Gon99}. For the purpose of the present paper it is sufficient
to know that spinors $\Phi_j$ can be subject to
the normalization condition
$$\int\limits_0^\pi\,\int\limits_0^{2\pi}(|\Phi_{j1}|^2+|\Phi_{j2}|^2)
\sin\vartheta d\vartheta d\varphi=1\> , \eqno(8)$$
{\it i. e.}, they form an orthonormal basis in $L_2^2({\Bbb S}^2)$.

The energy spectrum $\varepsilon$ of a quarkonium is given (in a more
symmetrical form than in ref. \cite{Gon01}) by the
relation $\varepsilon=\omega_1+\omega_2+\omega_3$ with
$$\omega_1=\omega_1(n_1,l_1,\lambda_1)=
\frac{-\Lambda_1 g^2a_1b_1+(n_1+\alpha_1)
\sqrt{(n_1^2+2n_1\alpha_1+\Lambda_1^2)\mu_0^2+g^2b_1^2(n_1^2+2n_1\alpha_1)}}
{n_1^2+2n_1\alpha_1+\Lambda_1^2}\>,\eqno(9)$$
$$\omega_2=\omega_2(n_2,l_2,\lambda_2)= $$
$$\frac{-\Lambda_2 g^2(a_1+a_2)(b_1+b_2)-(n_2+\alpha_2)
\sqrt{(n_2^2+2n_2\alpha_2+\Lambda_2^2)\mu_0^2+g^2(b_1+b_2)^2
(n_2^2+2n_2\alpha_2)}}
{n_2^2+2n_2\alpha_2+\Lambda_2^2}\>,\eqno(10)$$
$$\omega_3=\omega_3(n_3,l_3,\lambda_3)=
\frac{-\Lambda_3 g^2a_2b_2+(n_3+\alpha_3)
\sqrt{(n_3^2+2n_3\alpha_3+\Lambda_3^2)\mu_0^2+g^2b_2^2(n_3^2+2n_3\alpha_3)}}
{n_3^2+2n_3\alpha_3+\Lambda_3^2}\>,\eqno(11)$$
where
$\Lambda_1=\lambda_1-gB_1\>,\Lambda_2=\lambda_2+g(B_1+B_2)\>,
\Lambda_3=\lambda_3-gB_2\>,$
$n_j=0,1,2,...$, while $\lambda_j=\pm(l_j+1)$ are
the eigenvalues of euclidean Dirac operator
on unit sphere with $l_j=0,1,2,...$ Besides,
$$\alpha_1=\sqrt{\Lambda_1^2-g^2a_1^2}\>,
\alpha_2=\sqrt{\Lambda_2^2-g^2(a_1+a_2)^2}\>,
\alpha_3=\sqrt{\Lambda_3^2-g^2a_2^2}\>.\eqno(12)$$

Further, the radial part of (7), for instance, for the $\psi_1$-component, is
given at $n_1=0$ by
$$F_{11}=C_1Ar^{\alpha_1}e^{-\beta_1r}\left(1-
\frac{Y_1}{Z_1}\right),F_{12}=iC_1Br^{\alpha_1}e^{-\beta_1r}\left(1+
\frac{Y_1}{Z_1}\right),\eqno(13)$$
while at $n_1>0$ by
$$F_{11}=C_1Ar^{\alpha_1}e^{-\beta_1r}\left[\left(1-
\frac{Y_1}{Z_1}\right)L^{2\alpha_1}_{n_1}(r_1)+
\frac{AB}{Z_1}r_1L^{2\alpha_1+1}_{n_1-1}(r_1)\right],$$
$$F_{12}=iC_1Br^{\alpha_1}e^{-\beta_1r}\left[\left(1+
\frac{Y_1}{Z_1}\right)L^{2\alpha_1}_{n_1}(r_1)-
\frac{AB}{Z_1}r_1L^{2\alpha_1+1}_{n_1-1}(r_1)\right],\eqno(14)$$
with the Laguerre polynomials $L^\rho_{n_1}(r_1)$, $r_1=2\beta_1r$,
$\beta_1=\sqrt{\mu_0^2-(\omega_1-gA_1)^2+g^2b_1^2}$,
$A=gb_1+\beta_1$, $B=\mu_0+\omega_1-gA_1$,
$Y_1=[\alpha_1\beta_1- ga_1(\omega_1-gA_1)+g\alpha_1b_1]B+ g^2a_1b_1A$,
$Z_1=[(\lambda_1-gB_1)A+ga_1\mu_0)]B+ g^2a_1b_1A$.
Finally, $C_1$ is determined
from the normalization condition
$$\int_0^\infty(|F_{11}|^2+|F_{12}|^2)dr=\frac{1}{3}\>.\eqno(15)$$
Analogous relations will hold true for $\psi_{2,3}$, respectively,
by replacing
$a_1,A_1,b_1,B_1,\alpha_1 \to a_2,A_2,b_2,B_2,\alpha_3$ for $\psi_3$ and
$a_1,A_1,b_1,B_1,\alpha_1
\to -(a_1+a_2),-(A_1+A_2),-(b_1+b_2),-(B_1+B_2),\alpha_2$
for $\psi_2$ so that
$\beta_2=\sqrt{\mu_0^2-[\omega_2+g(A_1+A_2)]^2+g^2(b_1+b_2)^2}$,
$\beta_3=\sqrt{\mu_0^2-(\omega_3-gA_2)^2+g^2b_2^2}$.
Consequently, we shall gain that
$\psi_j\in L_2^{4}({\Bbb R}^3)$ at any $t\in{\Bbb R}$ and, as a result,
the solutions of (7) may describe relativistic bound states of a quarkonium
with the energy spectrum (9)--(11).

 Before apllying the above relations to a description of charmonium spectrum
let us adduce the nonrelativistic limits ({\it i. e.}, at $c\to\infty$) for
the energies of (9)--(11). The common case is not needed to us in the present
paper, so we shall restrict ourselves to the case of $n_j=0,1$ and $l_j=0$.
Expanding $\omega_j$ in $x=\frac{g}{\hbar c}$, we get
$$\omega_1(0,0,\lambda_1)=-x\frac{ga_1b_1}{\lambda_1}+\mu_0c^2\left[1-
\frac{1}{2}\left(\frac{a_1}{\lambda_1}\right)^2x^2+O(x^3)\right]\>,$$
$$\omega_1(1,0,\lambda_1)=-x\frac{ga_1b_1}{4\lambda_1}+
\mu_0c^2\left[1-
\frac{1}{8}\left(\frac{a_1}{\lambda_1}\right)^2x^2+O(x^3)\right]\>,\eqno(16)$$
which yields at $c\to\infty$ (putting $\hbar=c=1$ again)
$$\omega_1(0,0,\lambda_1)=
\mu_0\left[ 1-\frac{1}{2}\left(\frac{ga_1}{\lambda_1}\right)^2
\right]\>,
\omega_1(1,0,\lambda_1)=
\mu_0\left[ 1-\frac{1}{8}\left(\frac{ga_1}{\lambda_1}\right)^2
\right]\>.\eqno(17)$$
Analogously, we shall have
$$\omega_2(0,0,\lambda_2)=-\mu_0\left[ 1-\frac{1}{2}\left(\frac{g(a_1+a_2)}
{\lambda_2}\right)^2\right]\>,$$
$$\omega_2(1,0,\lambda_2)=-\mu_0\left[ 1-\frac{1}{8}\left(\frac{g(a_1+a_2)}
{\lambda_2}\right)^2\right]\>,\eqno(18)$$
$$\omega_3(0,0,\lambda_3)=
\mu_0\left[ 1-\frac{1}{2}\left(\frac{ga_2}{\lambda_3}\right)^2
\right]\>,
\omega_3(1,0,\lambda_3)=
\mu_0\left[ 1-\frac{1}{8}\left(\frac{ga_2}{\lambda_3}\right)^2
\right]\>,\eqno(19)$$
where, of course, $\lambda_j=\pm1$ and $\lambda_j^2=1$.

\section{Relativistic spectrum of charmonium}
  Now we can adduce numerical results for constants parametrizing
the charmonium spectrum which are shown in table I.

\begin{table}
\caption{Gauge coupling constant, mass parameter $\mu_0$ and
parameters of the confining SU(3)-connection for charmonium.}
\label{t.1}
\begin{center}
\begin{tabular}{cccccccc}
\noalign{\hrule}\\
$g$ & $\mu_0$ & $a_1$  & $a_2$ & $b_1$ & $b_2$ & $B_1$ &
$B_2$ \\
  & (GeV) &  &  & (GeV) & (GeV) & &  \\
\noalign{\hrule}\\
4.68010 & 0.627818 & 0.0516520 & 2.45565 & -0.705320 & 1.70660 &
2.10247 & 4.25862\\
\noalign{\hrule}\\
\end{tabular}
\end{center}
\end{table}

One can note that the obtained mass parameter $\mu_0$ is consistent with the
present-day experimental limits \cite{pdg} where the current mass of $c$-quark
($2\mu_0$) is accepted between 1.1 GeV and 1.4 GeV.
As for parameters $A_{1,2}$ of solution (5), only the wave functions depend on
them while the spectrum does not and within the present paper we consider 
$A_1=A_2=0$.

With the constants of Table I, the present-day levels of the charmonium 
spectrum were calculated with the help of (9)--(11) while their nonrelativistic
values with the aid of (17)--(19), according to the following combinations
(we use the notations of levels from ref.\cite{pdg}):
$$\eta_c(1S):
\varepsilon_1= \omega_1(0,0,-1)+\omega_2(0,0,-1)+\omega_3(0,0,-1)\>,$$
$$J/\psi(1S):
\varepsilon_2= \omega_1(0,0,-1)+\omega_2(0,0,1)+\omega_3(0,0,-1)\>,$$
$$\chi_{c0}(1P):
\varepsilon_3= \omega_1(0,0,-1)+\omega_2(0,0,-1)+\omega_3(0,0,1)\>,$$
$$\chi_{c1}(1P):
\varepsilon_4= \omega_1(0,0,1)+\omega_2(0,0,1)+\omega_3(0,0,1)\>,$$
$$\eta_{c}(1P):
\varepsilon_5= \omega_1(0,0,1)+\omega_2(1,0,-1)+\omega_3(1,0,-1)\>,$$
$$\chi_{c2}(1P):
\varepsilon_6= \omega_1(0,0,-1)+\omega_2(1,0,-1)+\omega_3(1,0,-1)\>,$$
$$\eta_c(2S):
\varepsilon_7= \omega_1(0,0,1)+\omega_2(1,0,1)+\omega_3(1,0,-1)\>,$$
$$\psi(2S):
\varepsilon_8= \omega_1(0,0,-1)+\omega_2(1,0,1)+\omega_3(1,0,-1)\>,$$
$$\psi(3770):
\varepsilon_9= \omega_1(1,0,-1)+\omega_2(1,0,-1)+\omega_3(0,0,1)\>,$$
$$\psi(4040):
\varepsilon_{10}= \omega_1(0,0,1)+\omega_2(0,0,-1)+\omega_3(1,0,-1)\>,$$
$$\psi(4160):
\varepsilon_{11}= \omega_1(0,0,1)+\omega_2(0,0,1)+\omega_3(1,0,-1)\>,$$
$$\psi(4415):
\varepsilon_{12}= \omega_1(0,0,1)+\omega_2(0,0,-1)+\omega_3(1,0,1)\>.
\eqno(20)$$

     Table II contains experimental values of these levels (from 
ref.\cite{pdg}) and our theoretical relativistic and nonrelativistic ones, and 
also the contribution of relativistic effects in \%, where it makes sense to
speak about such a contribution. Besides, one can notice that
the form of the wave functions (13)--(14) permits to consider, for instance,
the quantity $1/\beta_1$ to be a characteristic size of quarkonium state. 
Under the circumstances, if one calculates $1/\beta_1$ in both the relativistic 
($b_1\ne0$) and nonrelativistic ($b_1=0$) cases, then one can obtain those
sizes $r$ and $r_0$ in fm (1 fm = $10^{-13}$ cm) so the latter are adduced in
table II together with the quantity $r/r_0$.
\begin{table}
\caption{Experimental and theoretical charmonium levels.}
\label{t.2}
\begin{center}
\begin{tabular}{cccccccc}
\noalign{\hrule}\\
$\varepsilon_j$ &  Experim. & Relativ.  & Nonrelativ.
GeV & Relativ. contrib. & $r$ & $r_0$ & $r_0/r$\\
 & (GeV) & (GeV) & (GeV) & (\%) & (fm) & (fm) & \\
\noalign{\hrule}\\
$\varepsilon_1$ & 2.97980  & 2.97980  & 2.37202 & 20.3965 & 0.0603473 
& 1.32755 & 21.9984\\
$\varepsilon_2$ & 3.09688  & 3.09687 & 2.37202 & 23.4060 & 0.0603473
& 1.32755 & 21.9984\\
$\varepsilon_3$ & 3.41730  & 3.41729 & 2.37202 & 30.5877 & 0.0603473
& 1.32755 & 21.9984 \\
$\varepsilon_4$ & 3.51053  & 3.51764 & 2.37202 & 32.5678 & 0.0602972
& 1.32755 & 22.0167  \\
$\varepsilon_5$ & 3.52614  & 3.53085 & 1.05011 & 70.2590 & 0.0602972 
& 1.32755 & 22.0167  \\
$\varepsilon_6$ & 3.55617  & 3.54759 & 1.05011 & 70.3993 & 0.0603473
& 1.32755 & 21.9984\\
$\varepsilon_7$ & 3.59400  & 3.65282 & 1.05011 & 71.2521 & 0.0602972
& 1.32755 & 22.0167 \\
$\varepsilon_8$ & 3.68600  & 3.66955 & 1.05011 & 71.3831 & 0.0603473
& 1.32755 & 21.9984\\
$\varepsilon_9$ & 3.76990  & 3.75160 & -30.0324 &  & 0.0655399
& 2.64046 & 40.2878\\
$\varepsilon_{10}$ & 4.04000  & 4.04906 & 33.4683 &  & 0.0602972
& 1.32755 & 22.0167 \\
$\varepsilon_{11}$ & 4.16000  & 4.16614 & 33.4683 &   & 0.0602972 
& 1.32755 & 22.0167\\
$\varepsilon_{12}$ & 4.41500  & 4.41872 & 33.4683 &  & 0.0602972
& 1.32755 & 22.0167\\  
\noalign{\hrule}\\
\end{tabular}
\end{center}
\end{table}

\section{Physical interpretation}
     The results obtained allow us to draw a number of conclusions. As is seen
from table II, relativistic values are in good agreement with experimental
ones, while nonrelativistic ones are not. The contribution of relativistic
effects can amount to tens per cent and they cannot be considered as
small, as was expected by a number of theorists \cite{Du00}. Moreover, the
more excited the state of charmonium the worse the nonrelativistic 
approximation. For very excited states, the latter is not applicable at all.
The physical reason of it is quite clear. Really, we have seen in
the nonrelativistic limit (see the relations (16)--(19)) that the parameters
$b_{1,2}, B_{1,2}$ (see eq. (5)) of the linear interaction between quarks 
vanish under this limit and the
nonrelativistic spectrum is independent of them and is practically getting
the pure Coulomb one. As a consequence, the picture of linear confinement for
quarks should be considered as an essentially relativistic one while
the nonrelativistic limit is only a rather crude approximation. In fact, as 
follows from exact solutions of SU(3)-Yang--Mills equations of (5), the linear
interaction between quarks is connected with colour magnetic field that
dies out in the nonrelativistic limit, {\it i.e.} for static quarks. Only for
the moving rapidly enough quarks the above field will appear and generate
linear confinement between them. So the spectrum will depend on both
the static Coulomb colour electric field and the dynamical colour magnetic
field responsible for the linear confinement for quarks which is just confirmed
by our considerations. In our case, the interaction effect with the colour
magnetic field is taken into consideration from the very outset just reflects
the linear confinement at large distances.

Also, one can notice from table II that $r_0/r>>1$ for all the charmonium 
states, which additionally points out the importance of the relativistic 
effects connected with colour magnetic field for confinement.

Finally, I would like to say a few words concerning the nonrelativistic 
potential models often used in quarkonium theory. The potentials between 
quarks here are usually modelled by those of harmonic oscillator or of funnel 
type ({\it i. e.}, of the form $\alpha/r+\beta r$ with some constants $\alpha$ 
and $\beta$), see, {\it e. g.}, refs. \cite{{Rob},{Hag}}.

It is clear, however, that from the QCD point of view the interaction between
quarks should be described by the whole SU(3)-connection $A_\mu=A^c_\mu T_c$,
genuinely relativistic object, the nonrelativistic potential being only some
component of $A^c_t$ surviving in the nonrelativistic limit at $c\to\infty$.
As is easy to show, however, the connection of form $A^c_t=Br^\gamma$,
where $B$ is a constant, may be solution of the Yang-Mills equations (4) only
at $\gamma=-1$, {\it i. e.} in the Coulomb-like case. Consequently, 
the potentials employed in nonrelativistic approaches do not obey the Yang-Mills 
equations. The latter ones are essentially relativistic and, as we have seen, 
the components linear in $r$ of the whole $A_\mu$ are different from $A_t$ and
related with colour magnetic field vanishing in the nonrelativistic limit.
That is why the nonrelativistic potential approach seems to be inconsistent
though it developed a number of techniques and physical interpretations (in
particular in charmonium theory \cite{{Rob},{Hag}} which can be useful
under a relativistic description as well. Our approach uses only
the exact solutions of Yang-Mills equations as well
as in atomic physics the interaction among particles ({\it e. g.},
the electric Coulomb one) is always the exact solution of the Maxwell 
equations (the particular case of the Yang-Mills equations).
\section{Concluding Remarks}
As we have seen, the application of the Dirac equation to the charmonium
specrtum leads to a reasonable physical picture. From the obtained results
there follows that the standard approach of potential models on the base of
the Schr{\"o}dinger Equationwith some potential modelling confinement seems
to be inconsistent. The more consistent approach could be o the basis of the
Schr{\"o}dinger Equation in (colour) magnetic field since the linear 
confinement at large distances could be connected with the colour magnetic 
field rather than with the static colour electric one which follows from the
exact solution of the SU(3)-Yang-Mills equations (see ref. \cite{Gon01} and
the solution (5)). Historically, the latter way was rejected due to
incomprehensible reasons. The analysis of the present paper shows, however,
that the most consistent approach is probably the one based on the Dirac 
equation in the confining SU(3)-Yang-Mills field when the theory is 
relativistic from the very outset. In its turn, this approach is the direct
consequence of the {\it relativistic} QCD Lagrangian since the mentioned
Dirac equation is derived just from the latter one.

The calculations of the present paper can be extended. Indeed we have the
explicit form (7) for the relativistic wave functions of a quarkonium that
may be applied to the analysis of the quarkonium radiative decays. Besides,
our preliminary calculations show similar results to hold true also for
bottomonium. At last, there is a possibility of modifying the gluon propagator
on the basis of exact solutions of the SU(3)-Yang-Mills equations described
here and in ref. \cite{Gon01} for the mentioned propagator to be able to lead
to linear confinement between quarks at large ditances. The author hopes
to discuss the mentioned questions elsewhere.

\section{Acknowledgments}
This work was supported in part by the Russian
Foundation for Basic Research (grant No. 01-02-17157).

\end{document}